\documentclass{article}

\PassOptionsToPackage{numbers}{natbib}

\usepackage[preprint]{neurips_data_2024}

\usepackage[utf8]{inputenc} 
\usepackage[T1]{fontenc}    
\usepackage{hyperref}       
\usepackage{url}            
\usepackage{booktabs}       
\usepackage{amsfonts}       
\usepackage{nicefrac}       
\usepackage{microtype}      
\usepackage{xcolor}         

\usepackage{amsmath}

\usepackage{pifont}
\newcommand{\cmark}{\ding{51}}%
\newcommand{\xmark}{\ding{55}}%
\usepackage{bm}
\usepackage{mathrsfs}

\usepackage{algpseudocode}
\usepackage[linesnumbered,ruled,vlined]{algorithm2e}
\SetKwInOut{KwIn}{Input}
\SetKwInOut{KwOut}{Output}

\usepackage{graphicx}
\usepackage{subfig}

\usepackage{multirow}
\newcommand{\tabincell}[2]{\begin{tabular}{@{}#1@{}}#2\end{tabular}}
\usepackage{tabularx}
\usepackage{rotating}
\usepackage{adjustbox}
\usepackage{varwidth}
\usepackage{arydshln}
\usepackage{array}
\newcolumntype{P}[1]{>{\centering\arraybackslash}p{#1}}
\newcolumntype{M}[1]{>{\centering\arraybackslash}m{#1}}
\newcommand{\rowcentered}[1]{\begin{tabular}{l} #1 \end{tabular}}
\usepackage{makecell}

\usepackage{color}

\usepackage{enumitem}

\title{MatrixWorld: A Pursuit-Evasion Platform for Safe Multi-agent Coordination and Autocurricula}

\author{%
  Lijun Sun 
  \thanks{Department of Computer Science and Engineering, Southern University of Science and Technology, China} 
  \thanks{Faculty of Engineering and Information Technology, University of Technology Sydney, Australia} \\
  \texttt{11860004@mail.sustech.edu.cn} \\
  \And
  Yu-Cheng Chang $^{\dagger}$ \\
  \texttt{Yu-Cheng.Chang@uts.edu.au} \\
  \AND
  Chao Lyu 
  \thanks{College of Computer and Information Science, Southwest University, China} \\
  \texttt{lyuchao@swu.edu.cn} \\
  \And
  Chin-Teng Lin $^{\dagger}$ 
  \thanks{Corresponding author.} \\
  \texttt{Chin-Teng.Lin@uts.edu.au} \\
  \And
  Yuhui Shi $^{*\mathsection}$\\
  \texttt{shiyh@sustech.edu.cn} \\  
}

\begin{document}

\maketitle

\begin{abstract}
Multi-agent reinforcement learning (MARL) achieves encouraging performance in solving complex tasks.
However, the safety of MARL policies is one critical concern that impedes their real-world applications.
Popular multi-agent benchmarks focus on diverse tasks yet provide limited safety support.
Therefore, this work proposes a safety-constrained multi-agent environment: MatrixWorld\footnote{All codes are available at https://github.com/LijunSun90/MatrixWorld.}, based on the general pursuit-evasion game.
Particularly, a safety-constrained multi-agent action execution model is proposed for the software implementation of safe multi-agent environments based on diverse safety definitions.
It (1) extends the vertex conflict among homogeneous / cooperative agents to heterogeneous / adversarial settings, and (2) proposes three types of resolutions for each type of conflict, aiming at providing rational and unbiased feedback for safe MARL.
Besides, MatrixWorld is also a lightweight co-evolution framework for the learning of pursuit tasks, evasion tasks, or both, where more pursuit-evasion variants can be designed based on different practical meanings of safety.
As a brief survey, we review and analyze the co-evolution mechanism in the multi-agent setting, which clearly reveals its relationships with  autocurricula, self-play, arms races, and adversarial learning.
Thus, MatrixWorld can also serve as the first environment for autocurricula research, where ideas can be quickly verified and well understood.
\end{abstract}

\section{Introduction}
In thriving multi-agent reinforcement learning (MARL) research, increasingly complex multi-agent behaviors have emerged in increasingly complicated environments.
However, in current practice, MARL is still weak and challenging in guaranteeing the safety of multi-agent coordination \citep{zhang2021multi,oroojlooy2022review}, such as in autonomous driving \citep{wu2021flow}.
To date, the safe MARL has not gone beyond toy domains such as grid worlds.
Therefore, this paper revisits the significance of discrete pursuit-evasion game and introduces MatrixWorld, a new safety-constrained multi-agent pursuit-evasion platform.

Here, the safety concept is defined in terms of collision avoidance.
For instance, collisions may stem from conflicts of interest during multi-agent cooperation, the resolution of which determines the essential contribution of a coordination algorithm \citep{boutilier1996planning, boutilier1999sequential}.
When collisions occur, the multi-agent software environment should deal with these conflicts correctly by presenting reasonable and practical collision results and blaming the responsible one through e.g., rewards, based on which algorithms can learn properly.
However, as indicated by Terry et al. \citep{terry2021pettingzoo}, the real collision resolution mechanism implemented by MARL environments may be biased and lead to unexpected outcomes for the stochastic game (SG), where multi-agent actions should be executed simultaneously.
This may result from the bugs and complexities of codes when resolving the race conditions.

%

Popular multi-agent grid-world benchmarks provide limited safety support, based on which general MARL studies rarely report their safety performances even though they actually matter in associated applications. 
For example, in the Pursuit and Battle tasks of MAgent \citep{Zheng2018MAgent}, rewards are given in terms of the attacking action.
However, what if two agents (of the same team) intend to go to the same position?
The collision resolution mechanism is transparent without any reward feedback.
Similar situations apply	 to the predator-prey task of multi-agent particle environments (MPE) \citep{lowe2017multi}, the Pursuit task of PettingZoo \citep{terry2021pettingzoo}, etc.
In these cases, any biased or impractical collision resolution mechanisms with transparent information to the coordination algorithm will impede its safety learning.
Moreover, in practice, negative rewards for collisions are insufficient for guaranteeing the safety of MARL policies, and significant efforts are expected in open safe MARL research.

Another safety limitation of popular MARL grid-world environments is that their collision resolutions are not diverse enough to satisfy various safety requirements in applications.
This is because they focus more on providing diverse multi-agent behavior scenarios than diverse safety circumstances.
For example, in the same pursuit-evasion framework, if agents are drones, any collisions will lead both to crashes (death and disappearance).
However, if pursuers are predators and evaders are prey, predators will survive and prey will die after their collisions since they are unequal.
Furthermore, if pursuers are mobile agents while evaders are stationary targets, pursuers and evaders can collide while collisions between pursuers are not allowed, as in the multi-agent path finding (MAPF) problem.
These various safety definitions may bring different inequalities to the learning processes of agents and thus their resultant behaviors.

In addition, apart from the safety issue, we also consider the topic of autocurricula \citep{leibo2019autocurricula} in adversarial multi-agent research.
As a kind of automatic curriculum learning method, its related works include co-evolution, self-play, arms races, adversarial learning, etc.
These studies have achieved inspiring results in many fields, such as the genetic adversarial networks (GAN) \citep{goodfellow2014generative}, the   Grandmaster level agent in StarCraft II \citep{vinyals2019grandmaster},  agents with human-level performance in Quake III Arena \citep{jaderberg2019human}, and  emerged multi-agent strategic tool use and coordination in hide-and-seek \citep{baker2019emergent}.
However, many adversarial benchmarks are too heavy, making it computationally costly to test ideas and tune hyperparameters, such as the above video games.
Thus far, few studies have comprehensively discussed the relationships among these similar concepts, which may confuse new researchers in this field.

Motivated by the above concerns, we propose a  new safety-constrained multi-agent environment: MatrixWorld.
In particular, the pursuit-evasion game is selected as the basic environmental framework since it is general enough to be cooperative, competitive, or mixed.
Moreover, in the context of multi-agent co-evolution, a pursuit-evasion game may be the most intuitive framework due to its biological inspirations.
The main contributions of this work are as follows.

$\bullet$ For safe multi-agent coordination and MARL research, we propose a safety-constrained multi-agent action execution model to overcome limited safety support of popular MARL environments, 
To this end, we systematically analyze the agent-environment interaction models, the types of collisions and their resolutions in the adversarial multi-agent setting (Section \ref{sec_matrix_world}).

$\bullet$ For open autocurricula research, we present several pursuit-evasion variants as lightweight co-evolution frameworks.
As a brief survey, we review and analyze the co-evolution mechanism in the multi-agent setting, which clearly reveals its relationships with autocurricula, self-play, arms races, and adversarial learning (Section \ref{sec_coevolution}, \ref{sec_matrix_world}).


\section{Brief survey on co-evolution, autocurricula, and arms races}
\label{sec_coevolution}

\begin{figure*}[htbp!]
\centering
\includegraphics[width=0.7\linewidth]{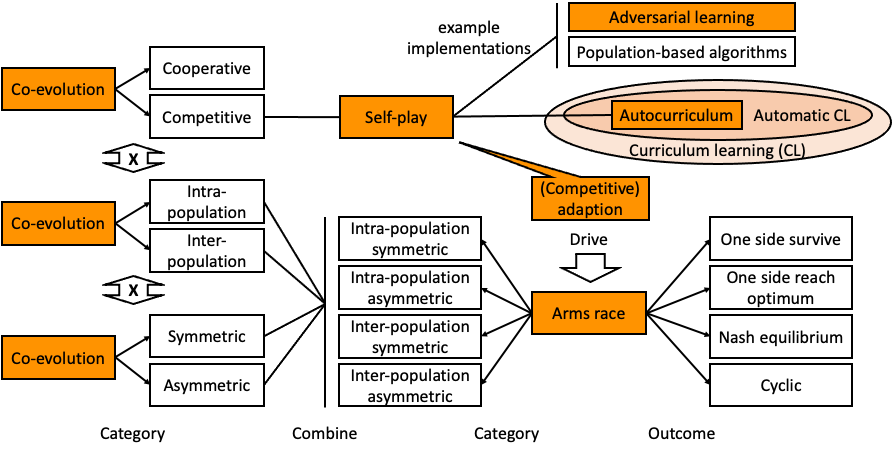}
\caption{
Relationships between co-evolution, self-play, autocurricula, arms races, and adversarial learning.
}
\label{fig_coevolution_mechanism}
\end{figure*}

In this section, we present a brief survey and clarify the relationships between several similar concepts: co-evolution, self-play, autocurricula, adaptation, arms races, and adversarial learning, as shown in Figure \ref{fig_coevolution_mechanism}.
We aim to better understand and highlight the role of co-evolution in multi-agent settings, more accurately use the related terminologies, and discuss some common misleading expectations and magics, especially for  researchers who are new to this field.
Finally, we propose that MatrixWorld can serve as the first environment for autocurriculum research, where ideas can be quickly verified and well understood.

\subsection{Co-evolution}
\label{sec_sub_coevolution}

Co-evolution is an natural phenomenon that witnesses the evolutionary progress and co-adaptations of, such as biological species.
As an evolutionary and learning mechanism, its appealing properties, including autonomous co-adaptation and arms races, the natural modeling of coupled relationships, and performance improvement, contribute to its sustained success in diverse research fields and applications.

The most distinct feature of co-evolution is that the fitness evaluation of one individual depends on other individuals, where each individual is a separately evolving or learning agent.
Co-evolution happens when adaptations occur between different agents, such as pursuers and evaders, environmental configurations and the agents therein.
Its effectiveness is promoted by iterative coordinated learning among diverse individuals, which may induce a steadier and more guaranteed improvement.
Therefore, co-evolution is more akin to a framework that can integrate different evolutionary computations (ECs), reinforcement learning  (RL) algorithms, etc.

The available co-evolution methods can be broadly classified into three categories.
First, there are competitive and cooperative co-evolution, which describe the relationships between individuals.
However, they do not determine the number of populations that co-evolve.
Individuals in one population can be competitive, while individuals from multiple populations may be cooperative.
The number of co-evolving populations defines the second category: intra-population and inter-population co-evolution methods.
Third, co-evolution can be classified into symmetric and asymmetric contests.
Symmetric co-evolution is for homogeneous agents in the sense that they learn similar skills for the same tasks.
In contrast, asymmetric co-evolution is for heterogeneous agents that target  different skills for different, perhaps competitive, tasks.

\subsection{Self-play}
\label{sec_self_play}

Self-play is a kind of competitive co-evolution process.
It generally refers to the competitive interactions of multiple agents but can also occur between different ``minds" (policy models) of the same physical agent, such as Alice and Bob in the research of Sukhbaatar et al. \citep{sukhbaatar2017intrinsic}.
Its idea is that learning progress is only achieved by the interplay or interaction between agents without external interference.
For example, Jaderberg et al. \citep{jaderberg2019human} trained symmetric adversarial agents in a competitive game (capture the flag) by letting the agents in a population play with each other.
Baker et al. \citep{baker2019emergent} trained asymmetric adversarial agents in another competitive game: hide-and-seek through multi-agent competitions.

\subsection{Adaptation and arms races}
\label{sec_arms_race}

Arms races between two rival sides are common in nature, such as those between predators and prey, and between parasites and hosts.
Biologically, Dawkins et al. \citep{dawkins1979arms} discussed an arm race as an evolutionary process of reciprocal counter-adaptations and the resultant challenges faced by two sides.
Based on this definition, the most crucial feature for identifying an arms race in a co-evolution process is the driving force derived from mutual counter-adaptations.
However, note that, adaptations and challenges do not necessarily mean more complex, stronger, or more general skills, which will be further explored in Section \ref{sec_autocurriculum}.

In addition, Dawkins et al. \citep{dawkins1979arms} proposed a two-way classification paradigms for arms races, i.e., symmetric or asymmetric, and interspecific or intraspcific, which form four possible combinations with biological examples.
This is also consistent with the categories of co-evolution.
Moreover, they summarized four classes of arms race endpoints: one side wins by driving the other to extinction, one side wins by first reaching its optimum, equilibrium endpoints, and cyclic endings.
One of the interesting discussions provided by Dawkins et al. \citep{dawkins1979arms} indicates that the average success of one side, such as the capture rate of predators, does not necessarily increase with the evolutionary time in an arms race.
This is because both rival sides are improving.
Then, the question of how to evaluate an arms race is progressing is actually the same as asking how the arms race will terminate, as described by the four possible outcomes of arms races.

\subsection{Curriculum learning (CL), automatic CL, and autocurricula}
\label{sec_autocurriculum}

Curriculum learning (CL) \citep{bengio2009curriculum} refers to a training strategy that learns from a sequence of related tasks organized with increasing difficulty, which generally achieves better results faster than normal learning methods.
Rather than manually designing a set of tasks and the time to switch between them, automatic CL automates part or all of the training process \citep{wang2021survey}.
In particular, the autocurriculum concept proposed by Leibo et al. \citep{leibo2019autocurricula} specifically refers to automatic CL in multi-agent settings, where the sequence of tasks or challenges is self-generated from mutual counter-adaptations in multi-agent interactions, i.e., an arms race.
In this sense, autocurriculum can be seen as equivalent to self-play.

However, based on both biological observations \citep{dawkins1979arms} and artificial intelligence (AI) investigations \citep{leibo2019autocurricula}, it is noted that challenges do not definitely mean more difficult or complex tasks, and adaptations to these challenges do not necessarily involve increasing strong abilities. 
Even the commonly adopted difficulty ranking of tasks, i.e., from easy to hard, in curriculum learning is not guaranteed to be optimal in all cases \citep{wang2021survey}.
The conditions for obtaining different arms race outcomes have attracted  researchers' attention.
Particularly, people are more interested in generating stronger models, such as agents with more complex behaviors.
Compared with various optimal rankings of task difficulties, the conclusions on how to avoid policy cycles are more consistent.
It has been shown that the policy cycling issue may occur in both symmetric and asymmetric arms races, for example, in a symmetric intra-population adversarial game of rock-paper-scissors \citep{leibo2019autocurricula} where the same agent strategy can play two different roles, and an asymmetric game \citep{nolfi1998co}.
Nolfi et al. \citep{nolfi1998co} investigated the policy cycling problem in the co-evolution and proposed that it could be reduced by evaluating the current policy with all past policies, which successfully drives the evolutionary process to an arms race with increasing complexity.
The same conclusion was also identified by Leibo et al. \citep{leibo2019autocurricula} and Vinyals et al. \citep{vinyals2019grandmaster}: self-play algorithms that do forgetting past policies provide a way to break out the policy cycling problem.
In addition, Leibo et al. \citep{leibo2019autocurricula} pointed out that the ceiling of the intelligence that can be achieved by an arms race is determined by the ``environment's carrying capacity", or the task definition that the autocuriculum is trying to solve.
Their research actually states that optima are defined by the problem itself.
For example, in a pursuit-evasion game without tools, novel tool usages will never emerge.

\subsection{Adversarial learning}
\label{sec_adversarial_learning}

Adversarial learning (AL) is a training framework that optimizes adversarial models with conflicting objectives and is typically implemented in an alternative learning mode.
Different from minimax search algorithms \citep{Russell2021Artificial} in which an agent optimizes itself against an optimal adversary, AL can be more general.
Subsequently, AL is not the only way to co-evolve adversarial agents and is more of an example implementation of self-play.
For example, in the symmetric adversarial game ``capture the flag" \citep{jaderberg2019human}, population-based RL was used to train a population of agents with similar skills that could play two adversarial game roles.

In addition, one main application of AL is to target an ultimate protagonist model rather than an adversary model, such as the generator model in generative adversarial networks (GANs) \citep{goodfellow2014generative} and  primary agents in the robust  adversarial reinforcement learning (RARL) \citep{pinto2017robust,ma2018improved,pan2019risk}.
Last, note that not all adversarial work refers to co-evolutionary adversarial learning, and they may be merely normal learning approaches in adversarial settings.
In these works, only one side in the adversarial setting is intentionally trained to be an opponent of the other, such as the adversarial policies in RL \citep{gleave2019adversarial} and adversarial examples.

\subsection{MatrixWorld: A lightweight co-evolution environment}
\label{sec_matrix_world_lightweight_coevolution}

Biologically, predators pursue their prey for dinner, while prey evade predators for life, and their co-evolutionary arms race and co-adaptations never stop.
In AI, the co-evolutionary pursuit-evasion has been studied since 1994 \citep{reynolds1994competition,miller1994protean,miller1994co,maes1996co}, which is slightly later than the research on pursuit or evasion alone, which has taken place since 1953 \citep{littlewood1953mathematician}.
Compared with complex 3D first-person video games such as Quake III Arena \citep{jaderberg2019human}, real-time strategy games such as StarCraft II \citep{samvelyan2019starcraft}, or adversarial games that need millions of episodes for arms races \citep{baker2019emergent}, MatrixWorld is a lightweight and simple co-evolution framework based on the pursuit-evasion game.

As discussed above, research on co-evolution is still open and it is supposed to be promising for automatically generating more complex agent behaviors and intelligence from the arms races in multi-agent interactions.
We propose that MatrixWorld can serve as the first environment for quickly verifying and effectively understanding research ideas in autocurricula.

\begin{table*}[htbp!]
\begin{center}
\scriptsize
\setlength
\tabcolsep{2pt}
\caption{Multi-agent-environment interaction models for distributed adversarial multi-agent systems.
Task environment properties \citep{Russell2021Artificial}:
\textbf{Single-agent vs. multi-agent}: whether the other agent(s) optimizes  some objectives that depend on the current agent.
\textbf{Cooperative vs. non-cooperative}: whether all agents share a common goal.
\textbf{Static vs. dynamic}: whether the environmental state or the agent's observation changes if the agent does nothing regardless of the flying time.
\textbf{Deterministic vs. nondeterministic (stochastic)}: whether ``the next state $s'$ is completely determined by the current state $s$ and the action $a$ executed by the agent(s)", i.e., whether the transition function $P(s'|s, a)=1$ or not.}
\label{tbl_agent_env_interact_model}
\begin{tabular}{c|c|c|c|l}
\hline\hline
Game of
& \tabincell{c}{Two-swarm\\turn taking} 
& \tabincell{c}{Agent-agent\\turn taking} 
& Agent-environment interaction model
& Environment model \& properties
\\ \hline
\multicolumn{3}{c|}{\rowcentered{Single-agent system}}
& \rowcentered{\includegraphics[width=0.25\linewidth]{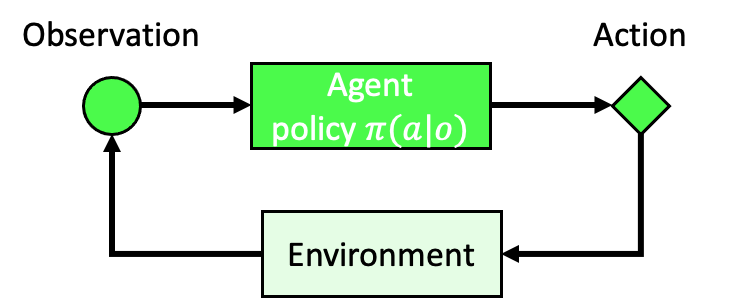}}
& \rowcentered{\tabincell{c}{
Markov decision process}}
\\ \hdashline
\multirow{3}{*}{\rotatebox[origin=c]{90}{Adversarial multi-agent system (e.g., pursuit-evasion)}}
& \rowcentered{\xmark}
& \rowcentered{\xmark}
& \rowcentered{\includegraphics[width=0.45\linewidth]{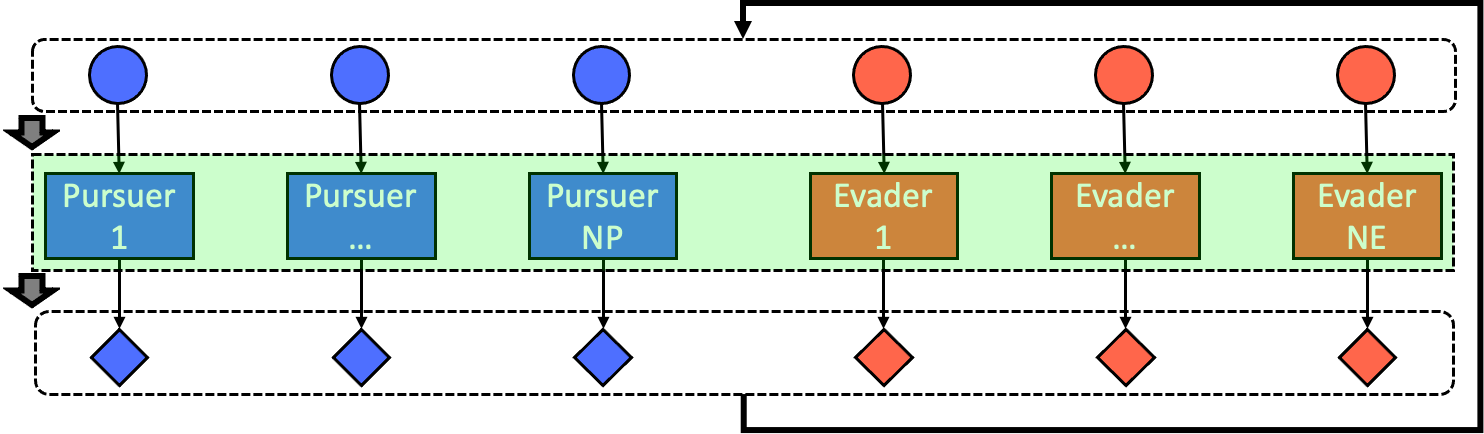}}
& \rowcentered{\tabincell{l}{
Stochastic / Markov game\\
- multi-agent\\
- non-cooperative\\
- dynamic\\
- stochastic}}
\\ \cdashline{2-5}
{}
& \rowcentered{\cmark}
& \rowcentered{\xmark}
& \rowcentered{\includegraphics[width=0.45\linewidth]{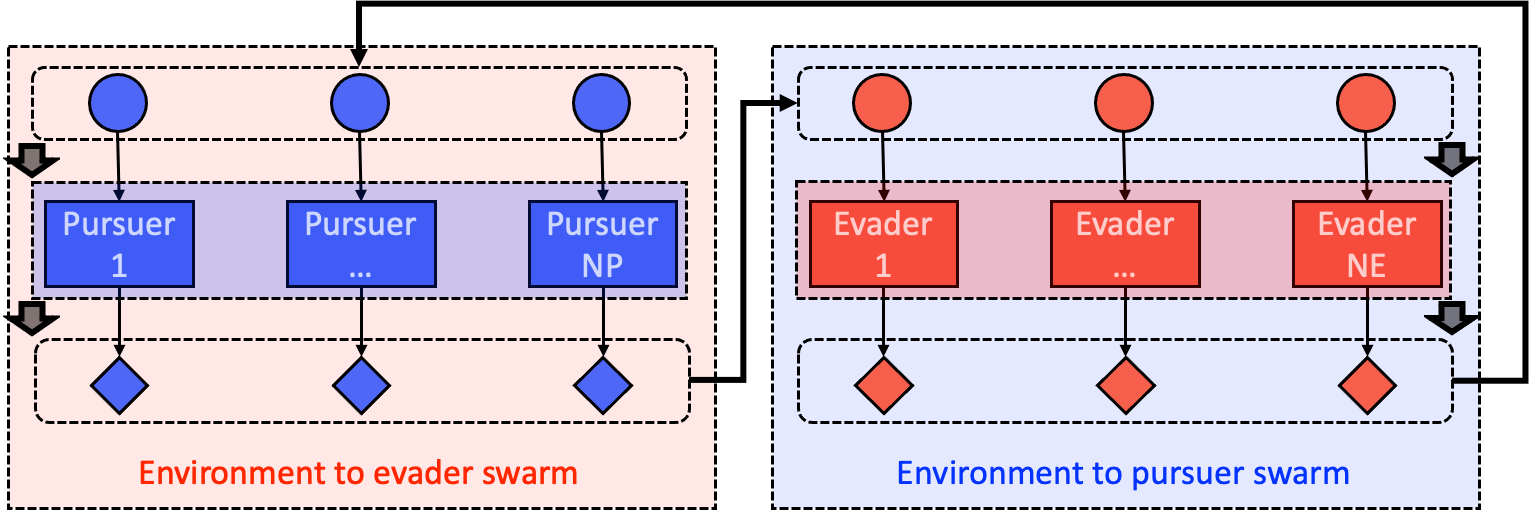}}
& \rowcentered{\tabincell{l}{
Two-swarm turn-taking game \\
- multi-agent\\
- non-cooperative\\
- dynamic\\
- stochastic}}
\\ \cdashline{2-5}
{}
& \rowcentered{\cmark}
& \rowcentered{\cmark}
& \rowcentered{\includegraphics[width=0.45\linewidth]{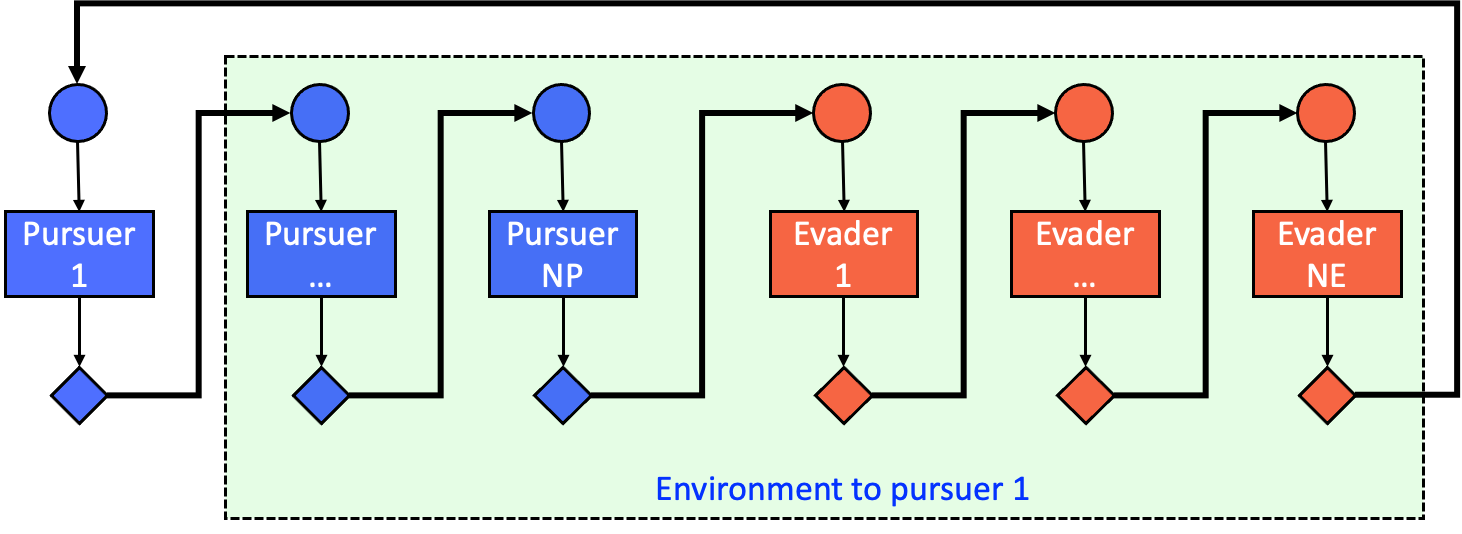}}
& \rowcentered{\tabincell{l}{
Agent environment cycle game\\
- multi-agent\\
- non-cooperative\\
- dynamic\\
- stochastic}}
\\ \hline
\end{tabular}
\end{center}
\end{table*}

\section{MatrixWorld: Safety-constrained multi-agent pursuit-evasion games}
\label{sec_matrix_world}

In this section, we introduce the propose MatrixWorld, a safety-constrained adversarial multi-agent environment based on co-evolutionary pursuit-evasion games.
In particular, we propose a safety-constrained multi-agent action execution model for the software implementation of safe multi-agent environments.
It determines the execution of agents' actions (the agent-environment interaction model in Section \ref{sec_agent_env_intect_model}), their results (the collision resolution mechanism in Section \ref{sec_collision_mechanism}), and who should be responsible, i.e., the evidence to give right rewards, etc.
In addition, we provide safety related information in the API for  safe MARL research.

\subsection{Multi-agent-environment interaction model}
\label{sec_agent_env_intect_model}

For a single-agent system, the typical agent-environment interaction (AEI) loop  is that: at each time step, (1) an agent receives an observation from the environment, (2) based on the observation the agent (policy) makes its decision and outputs an action, (3) the action is executed in the environment, and (4) the environment changes by responding to the action and other factors \citep{sutton2018reinforcement}.

For adversarial multi-agent settings, e.g., pursuit-evasion games, their agent-environment interaction models can be categorized based on two dimensions: whether agents take turns in a swarm-by-swarm manner, and whether agents take turns in an agent-by-agent manner, as shown in Table \ref{tbl_agent_env_interact_model}.
Therefore, the third row in the table is a strict stochastic game where all agents concurrently observe, make decisions, and execute actions. 
The fourth row in the table is a two-swarm turn-taking game and a one-swarm stochastic game for each swarm.
Similar to the two-player turn-taking game, a two-swarm turn-taking game can be defined as a game in which two swarms of agents take turns, i.e., swarm-by-swarm, observing, making decisions, and executing actions until the end of the game.
Finally, the last row in Table \ref{tbl_agent_env_interact_model} is a multi-agent turn-taking game, rather than a single-agent system, because the other agents also optimize some objectives in terms of the current agent in an AEI loop \citep{Russell2021Artificial}.

\subsection{Safety-constrained collision resolution mechanism}
\label{sec_collision_mechanism}

\begin{figure*}[htbp!]
\centering
\includegraphics[width=0.95\linewidth]{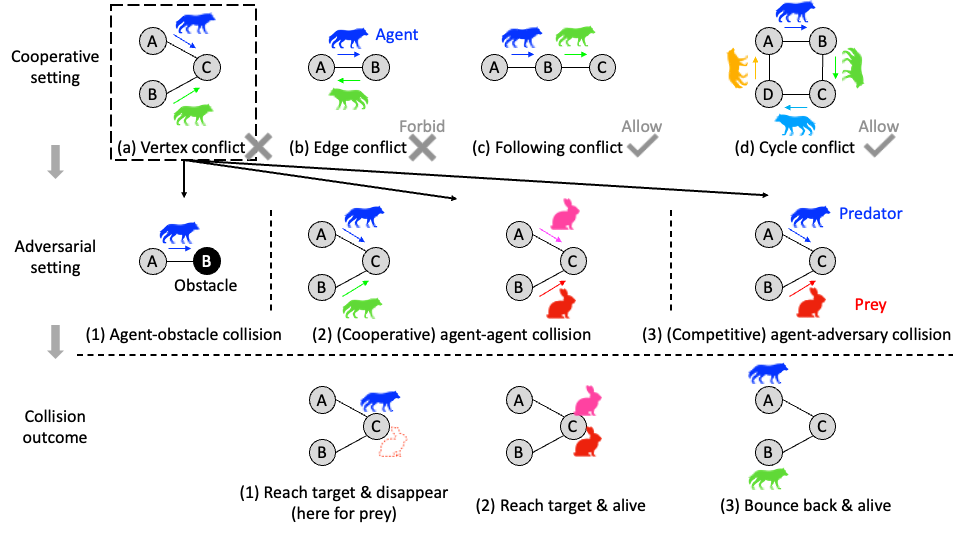}
\caption{Illustration of the types of collisions and outcomes in general multi-agent interactions.
First row: four types of conflicts for homogeneous / cooperative agents from \citep{gao2023review,stern2019multi}.
Second row: the extension of vertex conflict to heterogeneous / adversarial settings, i.e., the collision types.
Third row: the types of outcomes for each collision.
}
\label{fig_conflict_type}
\end{figure*}

There are four types of conflicts: (1) vertex conflict, (2) edge conflict, (3) following conflict, and (4) cycle conflict in MAPF \citep{gao2023review,stern2019multi}, as shown in the first row of Figure \ref{fig_conflict_type}.
Since the following and cycle conflicts do not actually cause collisions if all agents move simultaneously in a stochastic game, we allow these two ``conflicts" as the literature works.
Since the vertex and edge conflicts will physically cause the collision of agents, we forbid these two conflicts as the convention of MAPF.
However, the first difference is that the vertex conflict type in MAPF is for homogeneous / cooperative agents while our work extends it to heterogeneous / adversarial settings by further classifying it to three types of collisions based on the agent type (see the second row of Figure \ref{fig_conflict_type}).
The second difference is that rare works systematically summarize the environment execution outcome for each forbidden conflict type while our work proposes three types of resolutions based on the safety definition (see the third row of Figure \ref{fig_conflict_type}).

\begin{figure*}[htbp!]
\centering
\includegraphics[width=0.7\linewidth]{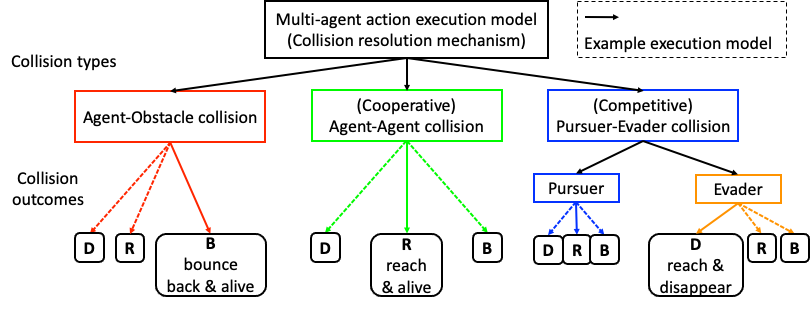}
\caption{Safety-constrained multi-agent collision resolution mechanism for the multi-agent environment modeled by stochastic game.}
\label{fig_action_execution_model}
\end{figure*}

In detail, we implement the safety constraints in MAS by designing a collision resolution mechanism based on meaningful and practical safety definitions.
As shown in Figure \ref{fig_conflict_type} and \ref{fig_action_execution_model}, collisions in the general multi-agent setting are classified into three types: (1) agent-obstacle collisions, (2) cooperative agent-agent collisions, and (3) competitive agent-adversary collisions.
Agent-obstacle collisions are common in single and multiple agent systems, where only one of the two colliding objects is the decision-maker that should be responsible.
Collisions between cooperative agents are not hostile and arise from conflicts of interest during cooperation, where both sides are accountable.
In contrast, competitive agent-adversary collisions may be intentional from one side and unexpected from the other, where the reward strategy depends on the given case.
Furthermore, we consider three outcomes for each collision: (1) reach (target) and disappear, (2) reach (target) and alive, and (3) bounce back (stay put) and alive.
These collision types, outcomes, and resultant reward strategies form the collision resolution mechanism that satisfies the safety requirements of different scenarios in Table \ref{tbl_collision_mechanism}.

\begin{table}[htbp!]
\begin{center}
\small
\setlength
\tabcolsep{2pt}
\caption{Rationality and example applications of various collision outcomes in the multi-agent setting.
\textbf{D:} reach (target) and disappear. 
\textbf{R:} reach (target) and alive.
\textbf{B:} bounce back (stay put) and alive.
}
\label{tbl_collision_mechanism}
\begin{tabular}{c|c|c|c|c|l}
\hline\hline
\multirow{2}{*}{Collision type}
& \multirow{2}{*}{Agent type}
& \multicolumn{3}{c|}{Outcome}
& \multirow{2}{*}{\tabincell{c}{Rationality\\(example application)}}
\\ \cline{3-5}
{}
& {}
& D
& R
& B
& {}
\\ \hline 
\multirow{3}{*}{Agent-obstacle}
& \multirow{3}{*}{Agent}
& \cmark
& -
& -
& Vehicle crash.
\\ \cline{3-6}
{}
& 
& -
& \cmark
& -
& Allowed in some MARL environments.
\\ \cline{3-6}
{}
& 
& -
& -
& \cmark
& Practical in many real-world applications.
\\ \hline
\multirow{3}{*}{Agent-agent}
& \multirow{3}{*}{Cooperative}
& \cmark
& -
& -
& Vehicle crash.
\\ \cline{3-6}
{}
&
& -
& \cmark
& -
& Allowed in some MARL environments.
\\ \cline{3-6}
{}
& {}
& -
& -
& \cmark
& Bumper cars.
\\ \hline
\multirow{8}{*}{Agent-adversary}
& Evader
& \cmark
& -
& -
& Vehicle crash.
\\ 
{}
& Pursuer
& \cmark
& -
& -
& 
\\ \cline{2-6}
{}
& Evader
& \cmark
& -
& -
& Predator eats prey.
\\ 
{}
& Pursuer
& -
& \cmark
& -
& 
\\ \cline{2-6}
{}
& Evader
& -
& -
& \cmark
& Pursuer is stronger.
\\ 
{}
& Pursuer
& -
& \cmark
& -
& 
\\ \cline{2-6}
{}
& Evader
& -
& -
& \cmark
& Bumper cars.
\\ 
{}
& Pursuer
& -
& -
& \cmark
& 
\\ \hline
\end{tabular}
\end{center}
\end{table}

\subsection{Pursuit-evasion game variants}
\label{sec_pursuit_evasion_variants}

\begin{figure*}[hbp!]
\centering
\includegraphics[width=0.91\linewidth]{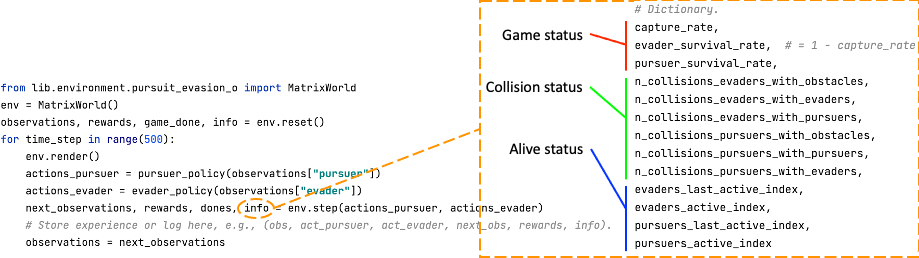}
\caption{Illustration of basic usage of MatrixWorld.}
\label{fig_api}
\end{figure*}

Pursuit-evasion game variants can be defined based on three dimensions: (1) the agent-environment interaction model (Section \ref{sec_agent_env_intect_model}), (2) the collision resolution mechanism (Section \ref{sec_collision_mechanism}), and (3) the capture behavior.
In view of the literature conventions and real-world applications, we propose nine pursuit-evasion tasks, i.e., the -D, -R, -B, -O, -S, -SB, -SD, -SDB, and -TO variants of the pursuit-evasion task in Table \ref{tbl_task_environment_definition}.

We mainly consider the first multi-agent-environment interaction model in Table \ref{tbl_agent_env_interact_model} for the stochastic game due to its generality in modeling multi-agent games, based on which most of the  pursuit-evasion variants are designed.
In addition, the last Pursuit-Evasion-TO task is based on the two-swarm turn-taking and agent-agent turn-taking model due to its advantages in theory analysis and the past sustained research \citep{bonato2011game}.

For cooperative agent-agent collisions, typical MARL tasks allow agents to collide with each other without fatal consequences, i.e., ``reach and alive".
Therefore, we include this in -R, -O, -S, and -SD variants of the Pursuit-Evasion task, which may ease the learning of complex strategies.
However, in some real-world applications, such collisions may be illegal, where the ``bounce back and alive" setting is actually applied.
Hence, the -B, -SB, and -SDB variants are designed accordingly to upgrade their safety.

For capture behaviors in the grid world, two main categories are considered.
One is the occupation-based capture in which an evader is captured if one or more pursuers occupy its position.
The other is the surrounding-based capture in which an evader is captured if enough pursuers and environment boundaries or obstacles encircle it such that it cannot move.

\begin{table*}[htbp!]
\begin{center}
\caption{\textbf{MatrixWorld}: Task definition.
Collision type: 
\textbf{A-O}: agent-obstacle;
\textbf{A-A}: cooperative agent-agent;
\textbf{P-E}: pursuer-evader.
Collision outcome:
\textbf{D}: reach (target) and disappear; 
\textbf{R}: reach (target) and alive;
\textbf{B}: bounce back (stay put) and alive.
}
\label{tbl_task_environment_definition}
\scriptsize
\setlength
\tabcolsep{2pt}
\begin{tabular}{c|c|c|c|c|c|c|c|c}
\Xhline{3\arrayrulewidth} 
\multicolumn{2}{c|}{\multirow{2}{*}{\tabincell{c}{Task environment\\(Pursuit-Evasion)}}}
& \multicolumn{2}{c|}{\tabincell{c}{Agent-environment\\interaction}} 
& \multicolumn{3}{c|}{\tabincell{c}{Collision\\resolution mechanism}} 
& \multirow{2}{*}{Capture behavior}
& \multirow{2}{*}{\tabincell{c}{Related work and applications\\(examples)}}
\\ \cline{3-7}
\multicolumn{2}{c|}{}
& \tabincell{c}{Two-swarm\\turn-taking}
& \tabincell{c}{Agent-agent\\turn-taking}
& A-O
& A-A
& P-E
& {}
\\ \Xhline{3\arrayrulewidth} 
\multirow{2}{*}{-D}
& Evader
& \multirow{2}{*}{\xmark}
& \multirow{2}{*}{\xmark}
& \multirow{2}{*}{D}
& \multirow{2}{*}{D}
& \multirow{2}{*}{D}
& \multirow{2}{*}{Occupation-based}
& \multirow{2}{*}{Real-world drone and vehicle swarm}
\\ \cline{2-2}
{}
& Pursuer
& {}
& {}
& {}
& {}
& {}
& {}
\\ \hline \hline
\multirow{2}{*}{-R}
& Evader
& \multirow{2}{*}{\xmark}
& \multirow{2}{*}{\xmark}
& \multirow{2}{*}{B}
& \multirow{2}{*}{R}
& \multirow{2}{*}{R}
& \multirow{2}{*}{Occupation-based}
& \multirow{2}{*}{\tabincell{c}{Multi-agent path finding (MAPF)\\MPE predator-prey \citep{lowe2017multi}}}
\\ \cline{2-2}
{}
& Pursuer
& {}
& {}
& {}
& {}
& {}
& {}
\\ \hline \hline
\multirow{2}{*}{-B}
& Evader
& \multirow{2}{*}{\xmark}
& \multirow{2}{*}{\xmark}
& \multirow{2}{*}{B}
& \multirow{2}{*}{B}
& \multirow{2}{*}{R}
& \multirow{2}{*}{Occupation-based}
& \multirow{2}{*}{Multi-agent path finding (MAPF)}
\\ \cline{2-2}
{}
& Pursuer
& {}
& {}
& {}
& {}
& {}
& {}
\\ \hline \hline
\multirow{2}{*}{-O}
& Evader
& \multirow{2}{*}{\xmark}
& \multirow{2}{*}{\xmark}
& \multirow{2}{*}{B}
& \multirow{2}{*}{R}
& D
& \multirow{2}{*}{Occupation-based}
& \multirow{2}{*}{Predator-prey; Spatial search}
\\ \cline{2-2}\cline{7-7}
{}
& Pursuer
& {}
& {}
& {}
& {}
& R
& {}
\\ \hline \hline
\multirow{2}{*}{-S}
& Evader
& \multirow{2}{*}{\xmark}
& \multirow{2}{*}{\xmark}
& \multirow{2}{*}{B}
& \multirow{2}{*}{R}
& B
& \multirow{2}{*}{Surrounding-based}
& \multirow{2}{*}{Benda et al. \citep{benda1986optimal}; Sun et al. \citep{sun2023toward}}
\\ \cline{2-2}\cline{7-7}
{}
& Pursuer
& {}
& {}
& {}
& {}
& R
& {}
\\ \hline \hline
\multirow{2}{*}{-SB}
& Evader
& \multirow{2}{*}{\xmark}
& \multirow{2}{*}{\xmark}
& \multirow{2}{*}{B}
& \multirow{2}{*}{B}
& B
& \multirow{2}{*}{Surrounding-based}
& \multirow{2}{*}{Predator-prey}
\\ \cline{2-2}\cline{7-7}
{}
& Pursuer
& {}
& {}
& {}
& {}
& R
& {}
\\ \hline \hline
\multirow{2}{*}{-SD}
& Evader
& \multirow{2}{*}{\xmark}
& \multirow{2}{*}{\xmark}
& \multirow{2}{*}{B}
& \multirow{2}{*}{R}
& B
& \multirow{2}{*}{\tabincell{c}{Surrounding-based\\disappear if captured}}
& \multirow{2}{*}{Predator-prey}
\\ \cline{2-2}\cline{7-7}
{}
& Pursuer
& {}
& {}
& {}
& {}
& R
& {}
\\ \hline \hline
\multirow{2}{*}{-SDB}
& Evader
& \multirow{2}{*}{\xmark}
& \multirow{2}{*}{\xmark}
& \multirow{2}{*}{B}
& \multirow{2}{*}{B}
& B
& \multirow{2}{*}{\tabincell{c}{Surrounding-based\\disappear if captured}}
& \multirow{2}{*}{Predator-prey}
\\ \cline{2-2}\cline{7-7}
{}
& Pursuer
& {}
& {}
& {}
& {}
& R
& {}
\\ \hline \hline
\multirow{2}{*}{-TO}
& Evader
& \multirow{2}{*}{\cmark}
& \multirow{2}{*}{\cmark}
& \multirow{2}{*}{B}
& \multirow{2}{*}{R}
& D
& \multirow{2}{*}{Occupation-based}
& \multirow{2}{*}{\tabincell{c}{Cops-and-robbers or vertex-pursuit \citep{quilliot1978jeux,NOWAKOWSKI1983235, bonato2011game};\\PettingZoo Pursuit \citep{terry2021pettingzoo}}}
\\ \cline{2-2}\cline{7-7}
{}
& Pursuer
& {}
& {}
& {}
& {}
& R
& {}
\\ \Xhline{3\arrayrulewidth} 
\end{tabular}
\end{center}
\end{table*}

\subsection{API}

The application programming interface (API) of MatrixWorld is designed based on the convention of RL community, as shown in Figure \ref{fig_api}.
The interface is the same while keeping the background multi-agent-environment interaction model in Table \ref{tbl_agent_env_interact_model} transparent for users' convenience.
In the dictionary information returned from the environmental step function, we provide (1) the game status data to monitor and compare the game progress regardless of the specific values in a reward structure; (2) the collision status data to give safety related feedback for performance evaluation, reward shaping, safety constraint construction, etc.; and (3) the ``alive" status data of the agents to track their remains and population decreases due to death.

\section{Conclusion}
\label{sec_conclusion}

In this paper, we propose a safety-constrained multi-agent benchmark: MatrixWorld.
Compared with popular MARL environments, it covers diverse safety definitions in various applications, by considering reasonable collision types, collision outcomes, and reward strategies in the collision resolution mechanism of multi-agent action execution model.
In addition, safety-related environmental information, such as different kinds of collisions in each time step, is provided in the API for conducting open safe MARL research, e.g., reward shaping.

Furthermore, MatrixWorld is also a lightweight co-evolution framework for  autocurriculum research based on the general pursuit-evasion game.
By revisiting the concepts of co-evolution, autocurricula, self-play, arms race, and adversarial learning, we clarify their relationships to achieve a better understanding and more accurate terminology usage.

Finally, this paper's work is only a start toward the safe MARL research.
For the rewarding mechanism, this paper's work only determines who should be responsible, but limited work is done on how to optimally utilize the safety related feedback and shape the rewards, which are crucial for the success of safe MARL algorithms.
Therefore, more work can be done based on the safety support of MatrixWorld and more multi-agent environments can be implemented based on the proposed general safety-constrained multi-agent action execution model.
It is also suggested to explore autocurricula in the lightweight MatrixWorld.

\section{Acknowledgements}
This work is partially supported by the Shenzhen Fundamental Research Program under Grant No. JCYJ20200109141235597, the National Science Foundation of China under Grant No. 61761136008, the Shenzhen Peacock Plan under Grant No. KQTD2016112514355531, the Program for Guangdong Introducing Innovative and Entrepreneurial Teams under Grant No. 2017ZT07X386, and  the Australian Research Council (ARC) under Discovery Grant DP210101093 and DP220100803.

\bibliographystyle{plainnat}
\bibliography{mybibfile}


\appendix


\section{Experiments}
\label{sec_experiments}

In this section, 
through adversarial learning, we achieve various arms race outcomes of different co-evolution mechanisms in MatrixWorld.
Based on experiments, arms races with steady and converging improvement are more practical for increasingly complex behaviors, while policy cycles between two rival sides are useful for producing diverse policies. 
In particular, we find that the passive (evasive) policy learning benefits more from co-evolution than active (pursuing) policy learning in an asymmetric adversarial game.
An arms race can drive the passive policy to a higher level than that in normal RL.
Finally, based on experiments in Pursuit-Evasion-S, we show the demanding safety assurance regarding the reward shaping or other techniques in safe MARL.

\subsection{Adversarial learning algorithms for pursuit-evasion tasks}
\label{sec_adversarial_learnng_algorithms}

We particularly explore three representative co-evolution frameworks and their influences on the arms race process and outcomes. 
\begin{itemize}
\item Pursuer specialist vs. evader specialist (Algorithm \ref{algorithm_specialist_vs_specialist}): Pursuers are trained to be specialized for the current evaders, and so are the evaders.
\item Pursuer generalist vs. evader specialist (Algorithm \ref{algorithm_generalist_vs_specialist}): Pursuers are trained to be general, while the evaders are trained to be specialized to their opponents, which is an unfair training system but may be more practical in the real world.
For example, police can learn from all past criminal data, while criminals may only access the current police data.
\item Pursuer generalist vs. evader generalist (Algorithm \ref{algorithm_generalist_vs_generalist}): Pursuers are trained to be general and robust for all past evaders, and so are the evaders.
\end{itemize}
The alternative learning scheme of adversarial learning is applied that when one side learns the other side is fixed.

\begin{algorithm}[htbp!]
\caption{Adversarial learning: Pursuer specialist vs. evader specialist}
\label{algorithm_specialist_vs_specialist}
Initialize pursuer model $\theta^p_0$ and evader model $\theta^e_0$. \\
\For {generation $k = 1 : N$}{
	$\theta^{p}_{k,k^p=0} = \theta^{p}_{k-1}$. \\
	\For {$k^p = 1 : N^p$}{
		Collect experiences by $\theta^{p}_{k,k^p - 1}$ and $\theta^{e}_{k - 1}$. \\
		Pursuer model update $\theta^{p}_{k,k^p - 1} \rightarrow \theta^{p}_{k,k^p}$.
	}
	$\theta^{p}_{k} = \theta^{p}_{k,N^p}$. \\
	$\theta^{e}_{k,k^e=0} = \theta^{e}_{k-1}$. \\
	\For {$k^e = 1 : N^e$}{
		Collect experiences by $\theta^{p}_{k}$ and $\theta^{e}_{k,k^e - 1}$. \\
		Evader model update $\theta^{e}_{k,k^e - 1} \rightarrow \theta^{e}_{k,k^e}$.
	}
	$\theta^{e}_{k} = \theta^{e}_{k,N^e}$.\\
\Return{$\theta^p_N, \theta^e_N$}.
}
\end{algorithm}
\begin{algorithm}[htbp!]
\caption{Adversarial learning: Pursuer generalist vs. evader specialist}
\label{algorithm_generalist_vs_specialist}
Initialize pursuer model $\theta^p_0$ and evader model $\theta^e_0$. \\
\For {generation $k = 1 : N$}{
	$\theta^{p}_{k,k^p=0} = \theta^{p}_{k-1}$. \\
	\For {$k^p = 1 : N^p$}{
		$k^e = k^p \bmod k$. \\
		Collect experiences by $\theta^{p}_{k,k^p - 1}$ and $\theta^{e}_{k^e}$. \\
		Pursuer model update $\theta^{p}_{k,k^p - 1} \rightarrow \theta^{p}_{k,k^p}$. \\
	}
	$\theta^{p}_{k} = \theta^{p}_{k,N^p}$. \\
	$\theta^{e}_{k,k^e=0} = \theta^{e}_{k-1}$. \\
	\For {$k^e = 1 : N^e$}{
		Collect experiences by $\theta^{p}_{k}$ and $\theta^{e}_{k,k^e - 1}$. \\
		Evader model update $\theta^{e}_{k,k^e - 1} \rightarrow \theta^{e}_{k,k^e}$.
	}
	$\theta^{e}_{k} = \theta^{e}_{k,N^e}$.\\
\Return{$\theta^p_N, \theta^e_N$}.
}
\end{algorithm}
\begin{algorithm}[htbp!]
\caption{Adversarial learning: Pursuer generalist vs. evader generalist}
\label{algorithm_generalist_vs_generalist}
Initialize pursuer model $\theta^p_0$ and evader model $\theta^e_0$. \\
\For {generation  $k = 1 : N$}{
	$\theta^{p}_{k,k^p=0} = \theta^{p}_{k-1}$. \\
	\For {$k^p = 1 : N^p$}{
		$k^e = k^p \bmod k$. \\
		Collect experiences by $\theta^{p}_{k,k^p - 1}$ and $\theta^{e}_{k^e}$. \\
		Pursuer model update $\theta^{p}_{k,k^p - 1} \rightarrow \theta^{p}_{k,k^p}$. \\
	}
	$\theta^{p}_{k} = \theta^{p}_{k,N^p}$. \\
	$\theta^{e}_{k,k^e=0} = \theta^{e}_{k-1}$. \\
	\For {$k^e = 1 : N^e$}{
		$k^p = k^e \bmod (k + 1)$. \\
		Collect experiences by $\theta^{p}_{k^p}$ and $\theta^{e}_{k,k^e - 1}$. \\
		Evader model update $\theta^{e}_{k,k^e - 1} \rightarrow \theta^{e}_{k,k^e}$. \\
	}
	$\theta^{e}_{k} = \theta^{e}_{k,N^e}$.\\
\Return{$\theta^p_N, \theta^e_N$}.
}
\end{algorithm}

\subsection{Experimental setup}
\label{sec_experimental_setup}

The policy models for the pursuers and evaders are both two-layer ReLU multi-layer perceptrons (MLPs) with hidden sizes of [400, 300].
The actor-critic \citep{sutton2018reinforcement} algorithm is adopted in the centralized training and decentralized learning (CTDE) scheme, with the policy and value learning rates set to $3\times10^{-4}$ and $10^{-3}$, respectively.
To stabilize the learning process, the actor model is trained 5 times, while value function is trained 1 time per epoch.
$N=30$ generations of co-evolution are conducted for Algorithms \ref{algorithm_specialist_vs_specialist}, \ref{algorithm_generalist_vs_specialist}, and \ref{algorithm_generalist_vs_generalist}, with $N_p = N_e = 400$ epochs in each generation.
For the Pursuit-Evasion-O task, 8 pursuers compete with 30 evaders in $40\times40$ grid worlds, 
while for Pursuit-Evasion-S, 20 pursuers compete with 5 evaders since 4 pursuers are required for every evader in the surrounding-based capture paradigm.
The reward functions are shown in Table \ref{tbl_reward function}.

\begin{table}[htbp!]
\begin{center}
\caption{Reward function for all pursuit-evasion tasks. ``–": the same.}
\label{tbl_reward function}
\small
\setlength
\tabcolsep{3pt}
\begin{tabular}{c|c|c|c}
\hline\hline
\multicolumn{2}{c|}{Pursuer}
& \multicolumn{2}{c}{Evader}
\\ \hline
Action
& Reward
& Action
& Reward
\\ \hline
Capture an evader
& 10
& Being captured
& -10
\\ \hline
Neighbor an evader
& 0.1
& Being neighbored
& -0.1
\\ \hline
Collide
& -12
& –
& -12
\\ \hline
Move before termination
& -0.05
& –
& -0.05
\\ \hline
\end{tabular}
\end{center}
\end{table}

\textbf{Suggestion:}
Set the hyperparameters $N_p$ and $N_e$ to allow the (moving average) learning performance improve to some extent.
In this way, an arms race can be observed in the adversarial learning; otherwise, neither rival side can learn effectively.

\subsection{Autocurricula in co-evolutionary pursuit-evasion tasks}
\label{sec_experiments_autocurriculum}

\begin{figure}[htbp!]
\centering
\subfloat{
\centering
\includegraphics[width=.75\linewidth]{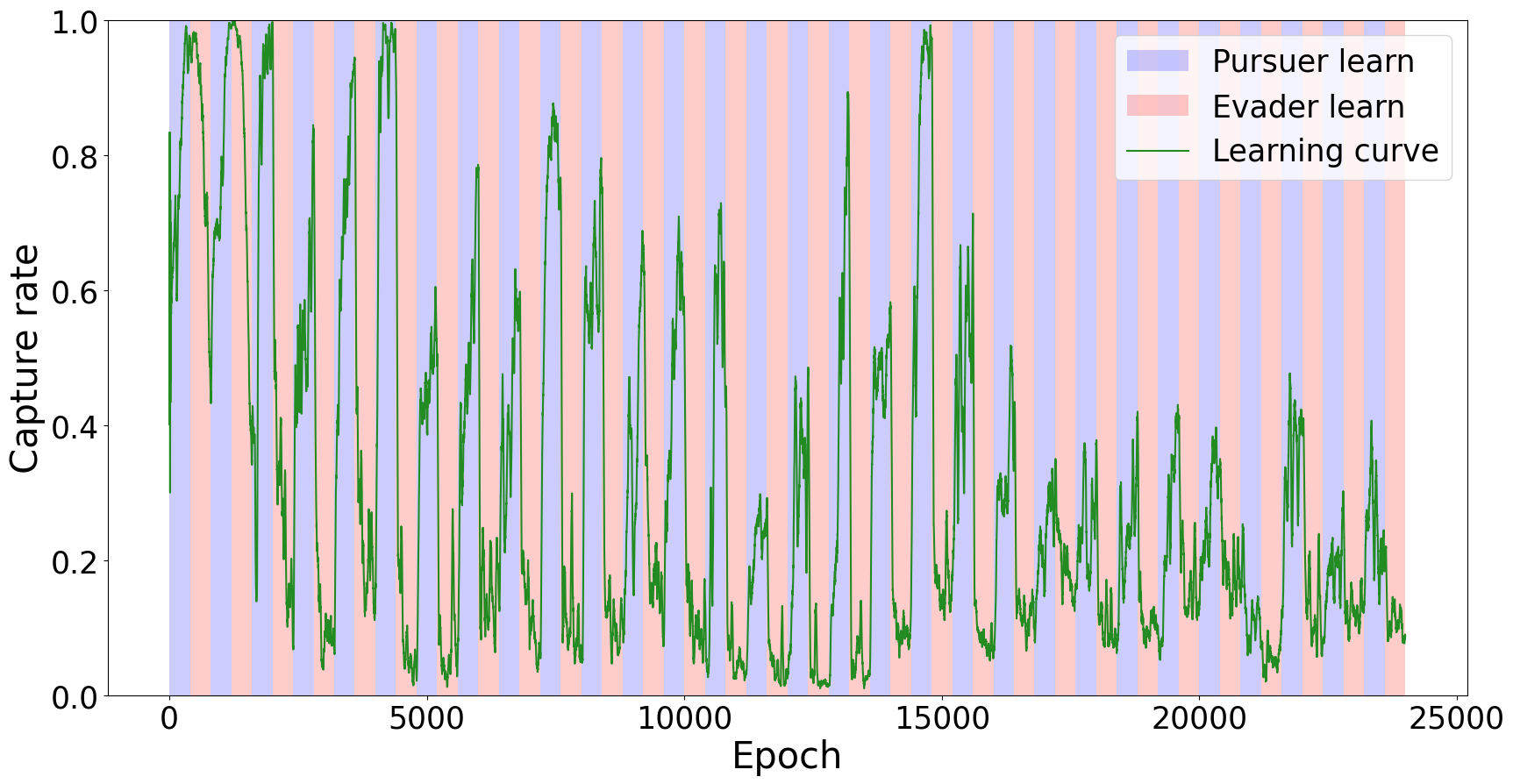}
\label{figure_capture_rate_specialist_vs_specialist}}
\\
\subfloat{
\centering
\includegraphics[width=.75\linewidth]{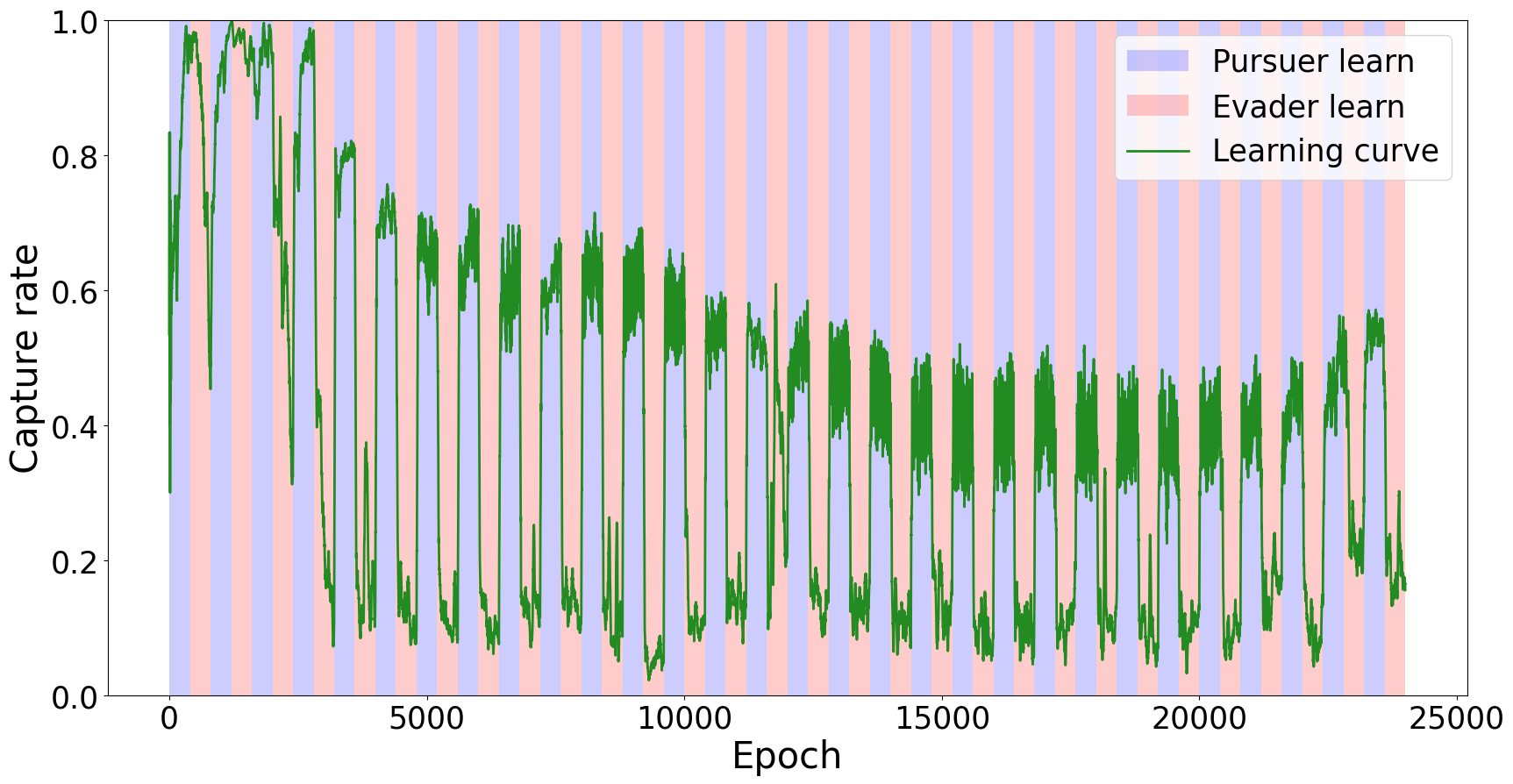}
\label{figure_capture_rate_generalist_vs_speicalist}}
\\
\subfloat{
\centering
\includegraphics[width=.75\linewidth]{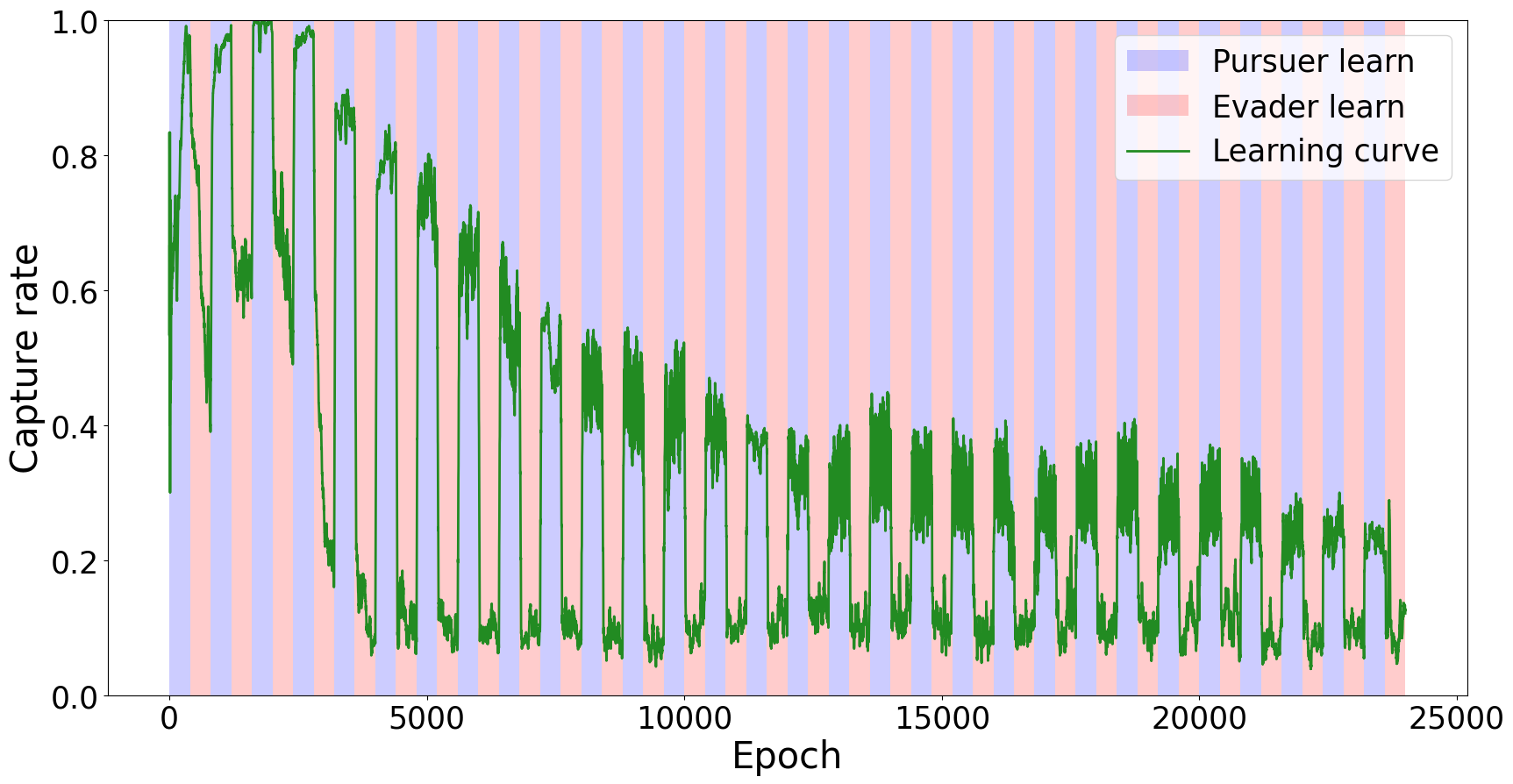}
\label{figure_capture_rate_generalist_vs_generalist}}
\caption{Training performance achieved for Pursuit-Evasion-O by Algorithms \ref{algorithm_specialist_vs_specialist}, \ref{algorithm_generalist_vs_specialist}, and \ref{algorithm_generalist_vs_generalist} (from top to bottom). 
The curves are smoothed over 30 points.}
\label{fig_capture_rate}
\end{figure}

\textbf{Adaptation and arms race occur between pursuers and evaders.}
As shown in the training performance presented in Figure \ref{fig_capture_rate}, the learning of pursuers brings challenges to the evaders' policies and the capture rate increases, and vice versa.
The performance of the agents (pursuers and evaders) continues to improve with iterative generations.
Their performance variance brought by policy adaptations tends to converge as the number of generations increases, but this is not the end of the arms race. 
When we observe many more generations than $N=30$, say 100 generations, we  find that the performance variance diverges again, which demonstrates the sustained adaptations during the arms race.

\begin{figure}[htbp!]
\centering
\subfloat[Pursuer specialist vs. evader specialist (Algorithm \ref{algorithm_specialist_vs_specialist})]{
\centering
\includegraphics[width=0.97\linewidth]{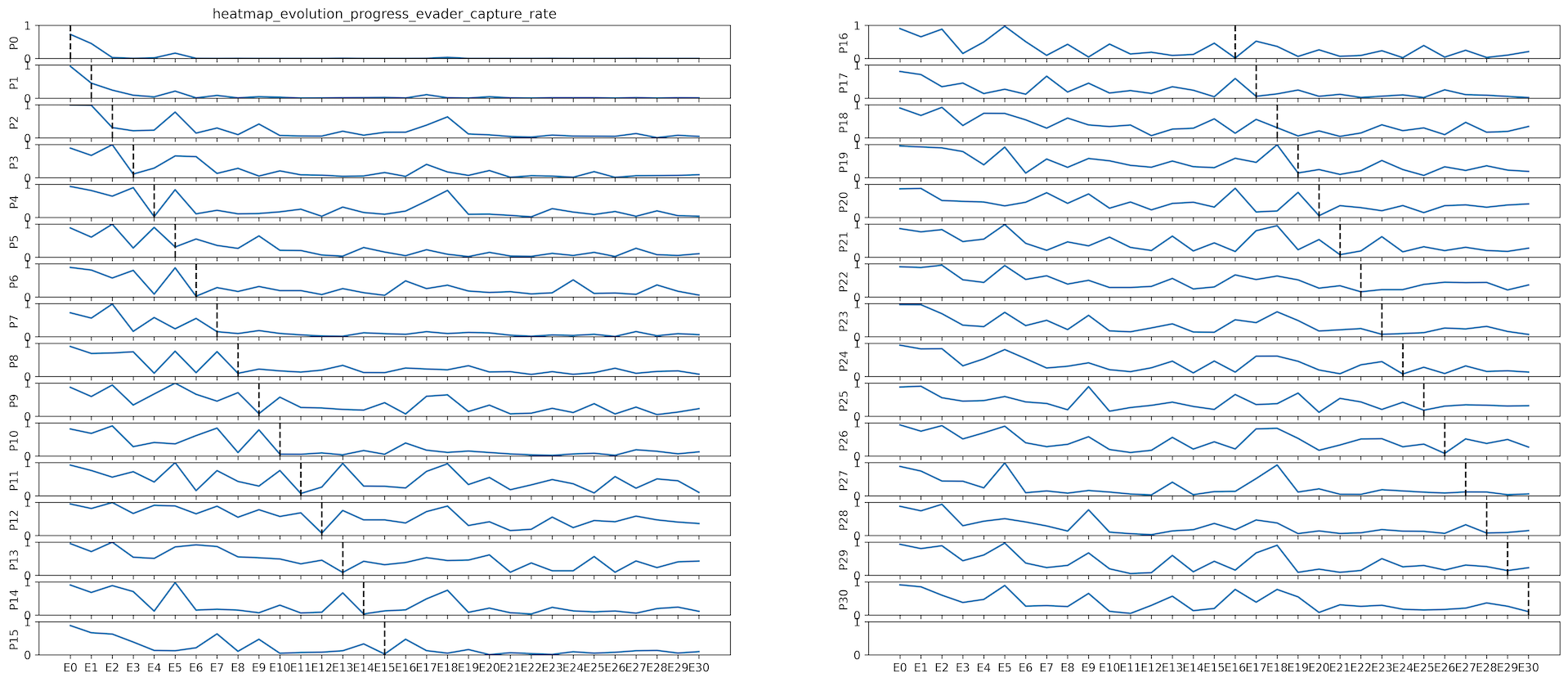}
\label{figure_evolution_progress_specialist_vs_specialist}}
\\
\subfloat[Pursuer generalist vs. evader specialist (Algorithm \ref{algorithm_generalist_vs_specialist})]{
\centering
\includegraphics[width=0.97\linewidth]{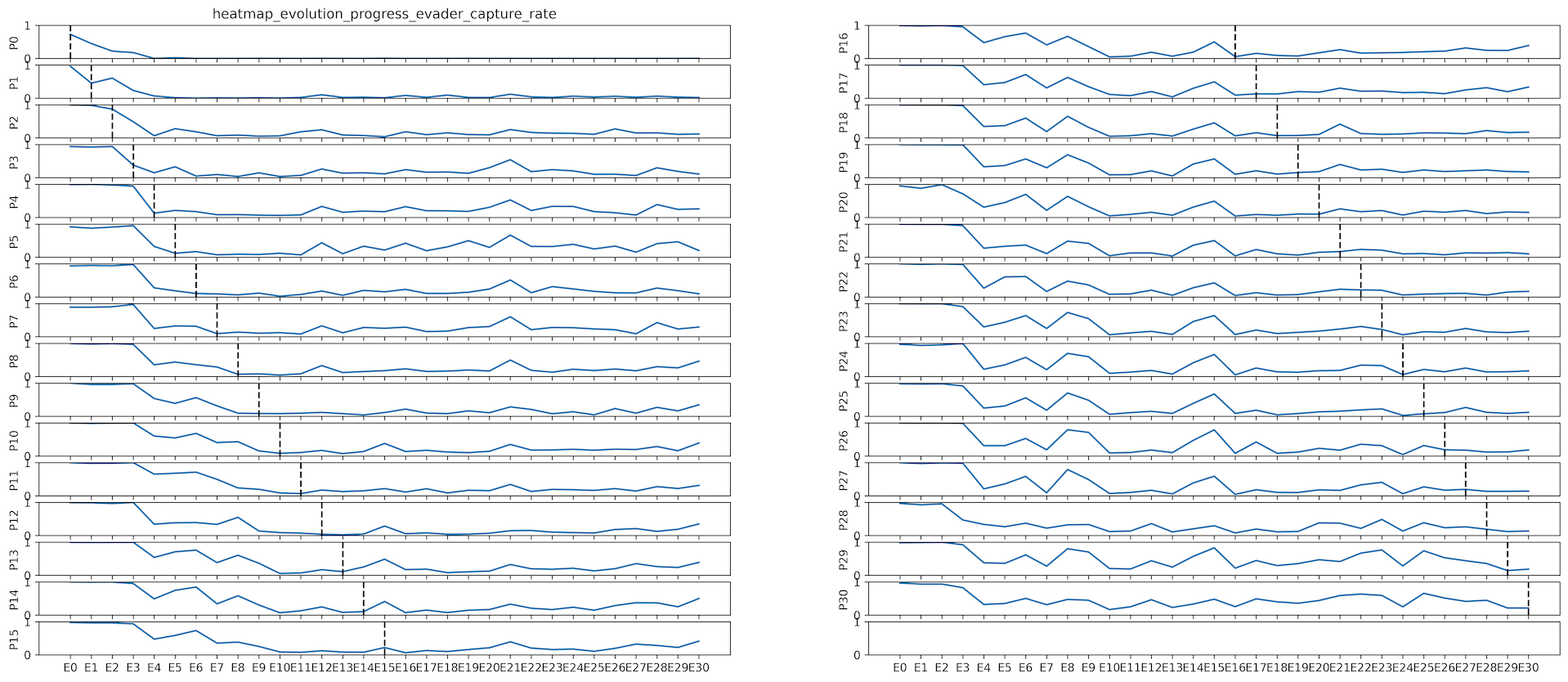}
\label{figure_evolution_progress_generalist_vs_specialist}}
\\
\subfloat[Pursuer generalist vs. evader generalist (Algorithm \ref{algorithm_generalist_vs_generalist})]{
\centering
\includegraphics[width=0.97\linewidth]{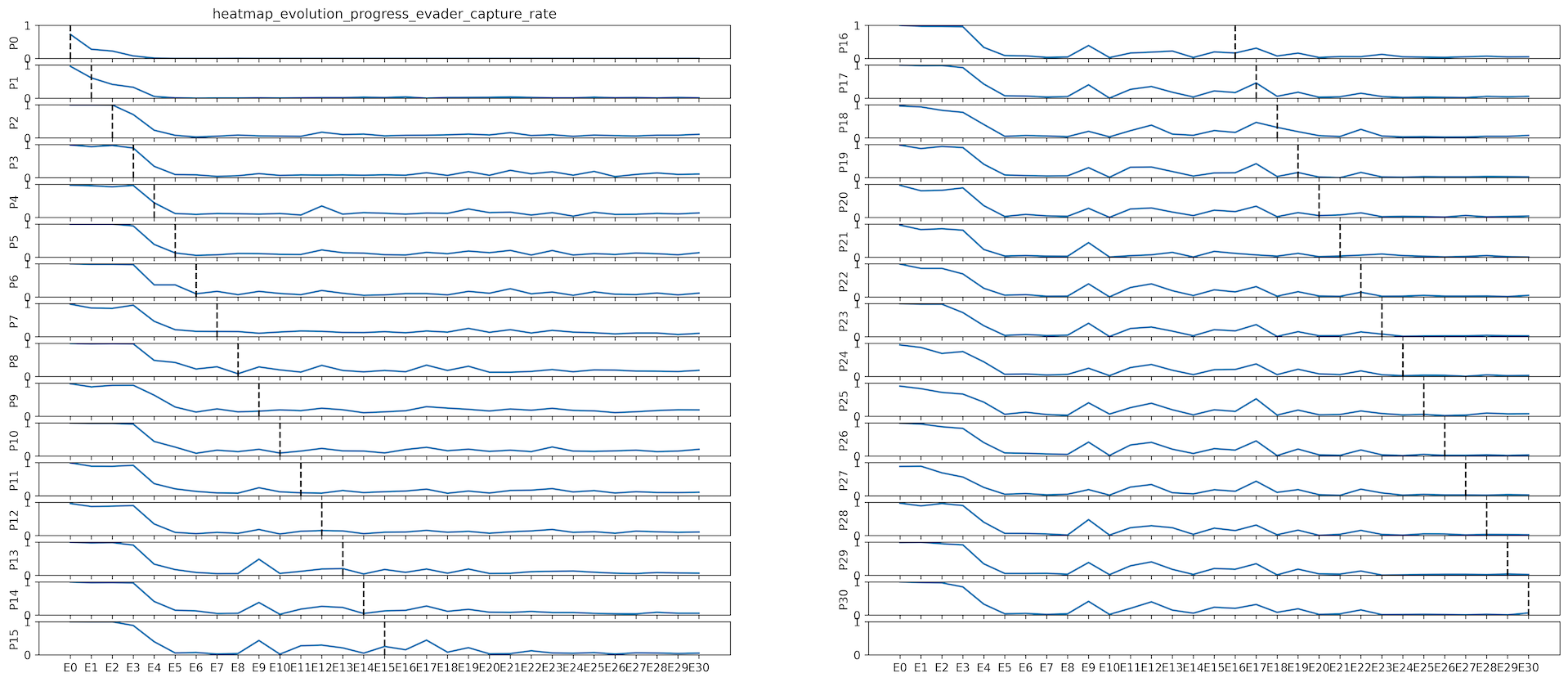}
\label{figure_evolution_progress_generalist_vs_generalist}}
\caption{Evolutionary performance (capture rate) of evaders based on the testing performance achieved on Pursuit-Evasion-O, which is averaged over 10 independent runs.
Horizontal axis: E0 - E30, the 30 generations of evaders. 
Vertical axis: P0 - P15 (left), P16 - P30 (right), the 30 generations of pursuers. E0 and P0 are the initial evaders and pursuers.
A black dashed line indicates the associated pursuer and evader are in the same generation.}
\label{fig_evolution_progress_capture_rate}
\end{figure}

\textbf{Policy cycles occur in the specialist-vs.-specialist framework, but they are avoided in the generalist-vs.-generalist scheme.}
As shown in Figure \ref{fig_evolution_progress_capture_rate}, we test the evolutionary performance (capture rate) of the evaders against the pursuers in each generation.
The policy cycles are observed in Figure \ref{figure_evolution_progress_specialist_vs_specialist}, where both the pursuers and evaders learn only against their contemporary opponents.
The evaders' performances against the pursuers in a  specific generation, i.e., one row in the figure, fluctuate periodically from generation to generation, or even decrease sometimes. 
This indicates that due to the mutual adaptations of pursuers and evaders, their skills periodically appear and disappear, which is called a policy cycle \citep{nolfi1998co}. 
On the other hand, such policy cycles from counter-adaptations are useful for producing diverse policies with similar complexity levels.
In contrast, when the pursuers are trained to be general in Algorithm \ref{algorithm_generalist_vs_specialist}, the policy cycle problem is less severe in Figure \ref{figure_evolution_progress_generalist_vs_specialist}.
Furthermore, when both the pursuers and evaders learn against all past opponents, the skill of competing with a specific pursuer is preserved in later generations, which can be seen from the stable performance in Figure \ref{figure_evolution_progress_generalist_vs_generalist} after the evaders first learn against that pursuer.
This indicates that learning against past opponents helps avoid forgetting already learned skills.
This result is consistent with the idea of past literature works on eliminating policy cycles, such as \citep{nolfi1998co,vinyals2019grandmaster,leibo2019autocurricula}.

\begin{figure}[htbp!]
\centering
\subfloat[Pursuer specialist vs. evader specialist (Algorithm \ref{algorithm_specialist_vs_specialist})]{
\centering
\includegraphics[width=0.6\linewidth]{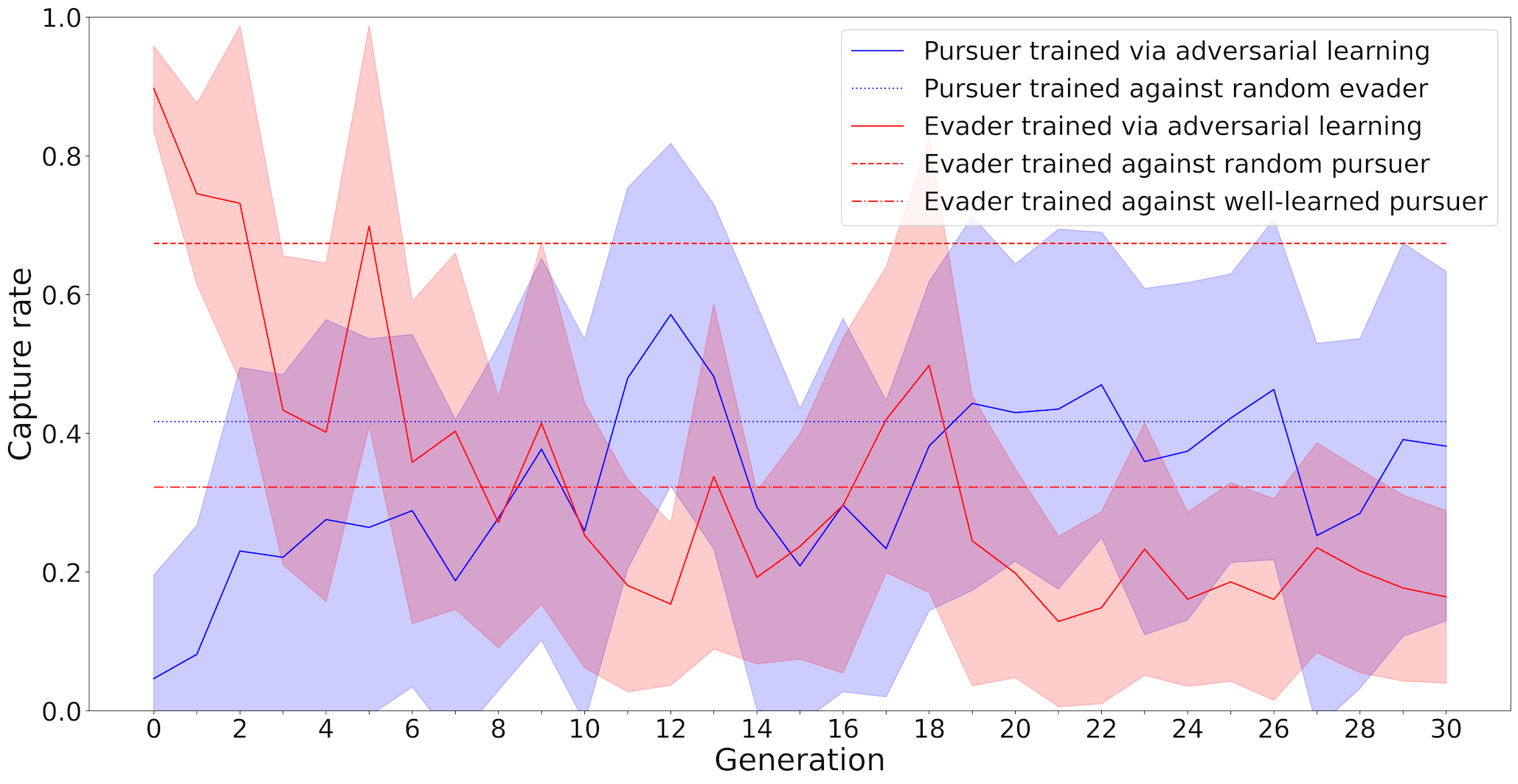}
\label{figure_generalization_performance_specialist_vs_specialist}}
\\
\subfloat[Pursuer generalist vs. evader specialist (Algorithm \ref{algorithm_generalist_vs_specialist})]{
\centering
\includegraphics[width=0.6\linewidth]{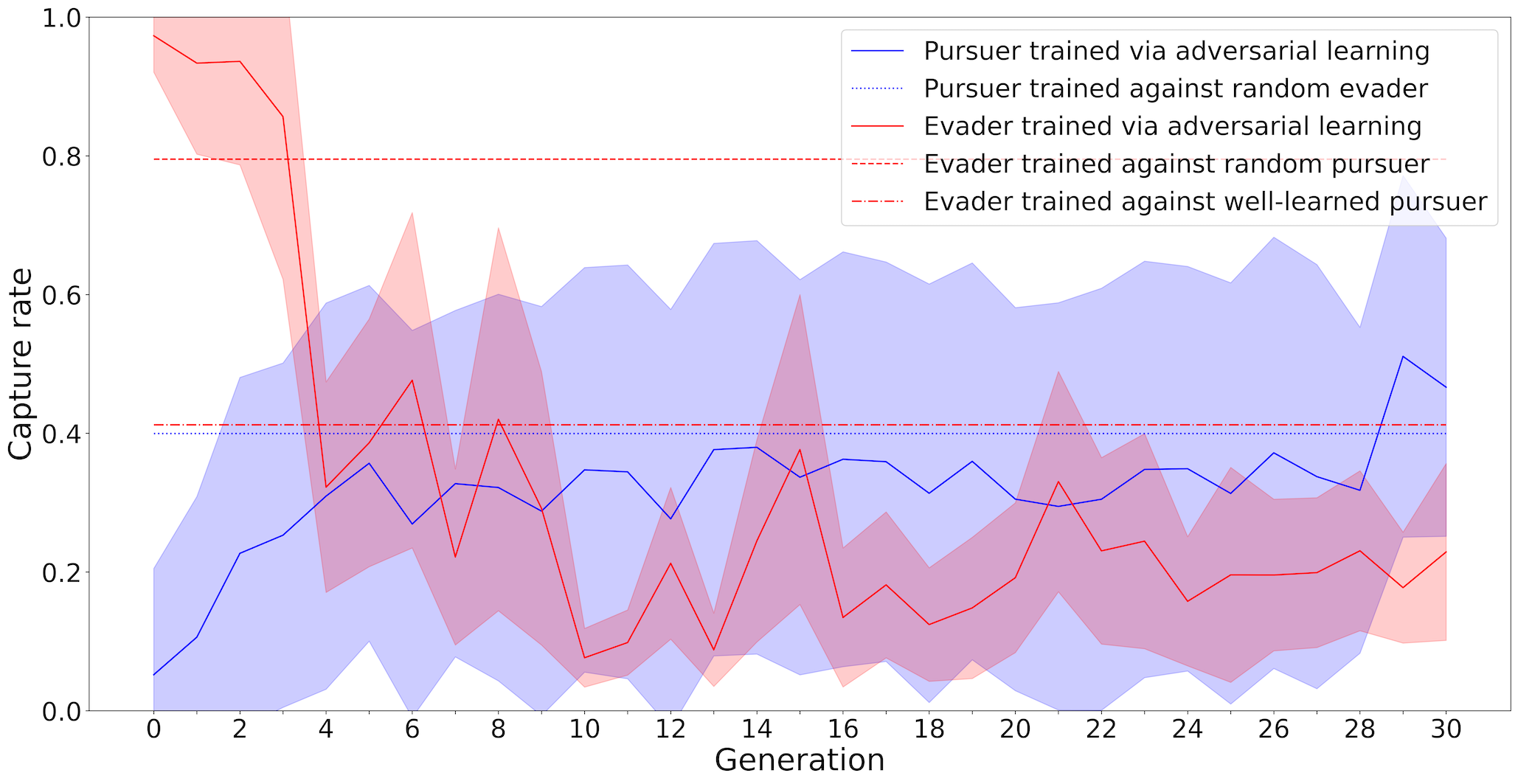}
\label{figure_generalization_performance_generalist_vs_speicalist}}
\\
\subfloat[Pursuer generalist vs. evader generalist (Algorithm \ref{algorithm_generalist_vs_generalist})]{
\centering
\includegraphics[width=0.6\linewidth]{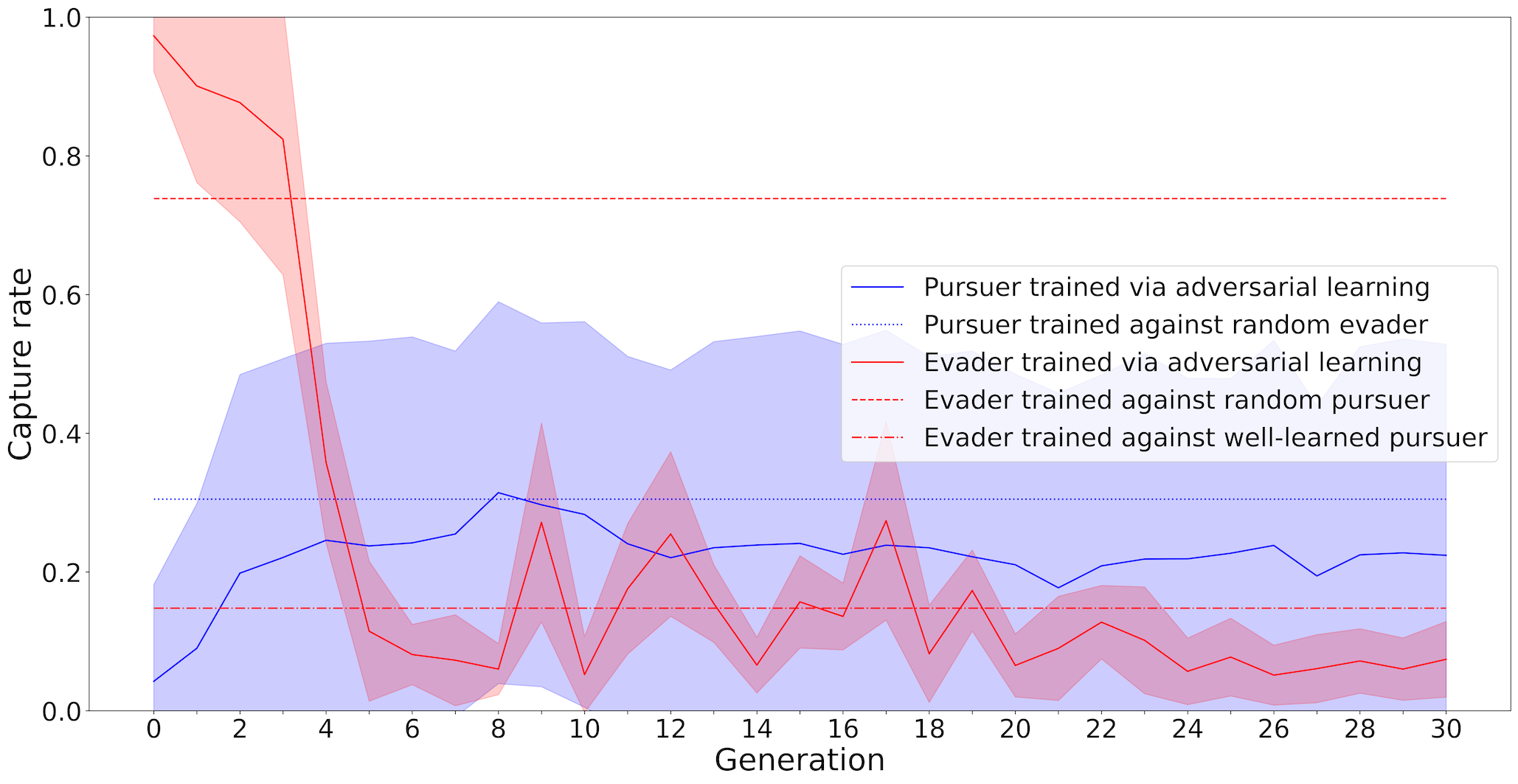}
\label{figure_generalization_performance_generalist_vs_generalist}}
\caption{Generalization performance achieved for Pursuit-Evasion-O.}
\label{fig_generalization_performance}
\end{figure}

\textbf{An arms race is more useful for learning passive policies.}
In Figure \ref{fig_generalization_performance}, we test the generalization performance of the agent in each generation by averaging its performance against the opponents across all generations.
At the same time, we compare it with three baselines.
\begin{itemize}
\item Baseline 1: the pursuer that is trained by competing with randomly walking evaders.
\item Baseline 2: the evader that is trained by competing with randomly walking pursuers.
\item Baseline 3: the evader that is trained by competing with well-learned pursuers. 
The well-learned pursuer is trained by competing with randomly walking evaders. 
That is, the baseline 3 is trained via manual curriculum learning.
\end{itemize}

The conclusion that an arms race is more useful for learning passive (evasive) policies can be drawn based on two consistent observations from Figure \ref{figure_generalization_performance_specialist_vs_specialist} to  \ref{figure_generalization_performance_generalist_vs_generalist}.
First, adversarial learned evaders have better generalization performance than the baseline evaders, while adversarial learned pursuers do not achieve significantly better generalization performance than the baseline pursuers.
Second, it can be seen that the complexity of a passive evasive policy highly depends on opponents' ability.
When trained only against random pursuers, the evaders’ performance is hard to improve once it reaches a certain level, which is the worst in Figure \ref{fig_generalization_performance}. 
If trained against stronger pursuers, i.e., well-learned pursuer, their performance is significantly enhanced. 
The best evader policies are obtained by adversarial learning. 
This indicates that autocurriculum learning is achieved and increasingly strong pursuers drive the continuous learning of evaders.

\subsection{General MARL in safe multi-agent coordination scenarios}
\label{sec_experiments_safe_MARL}

\begin{figure}[htbp!]  
\centering
\subfloat{
\centering
\includegraphics[width=0.6\linewidth]{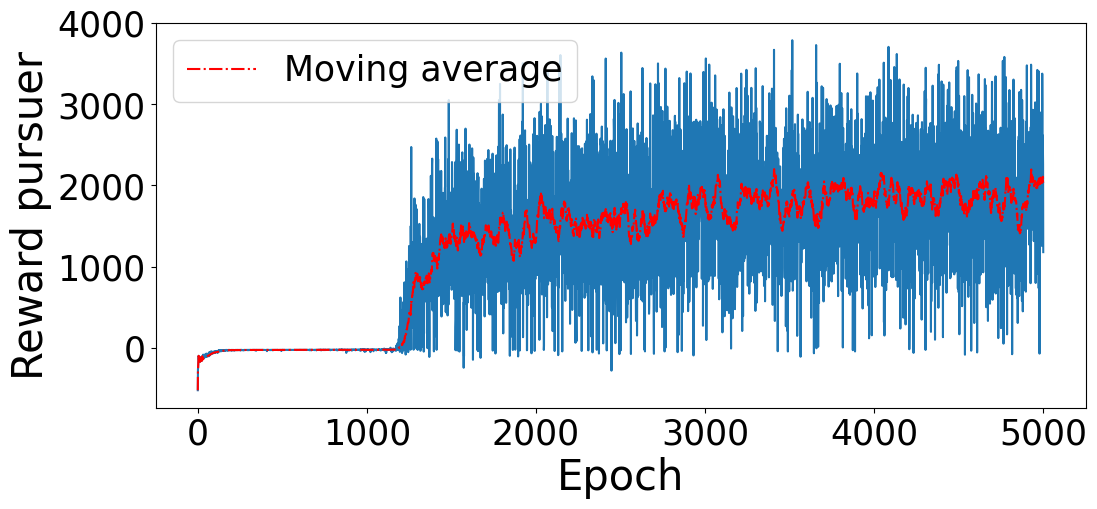}
\label{figure_pursuit_evasion_s_reward}}
\\
\subfloat{
\centering
\includegraphics[width=0.6\linewidth]{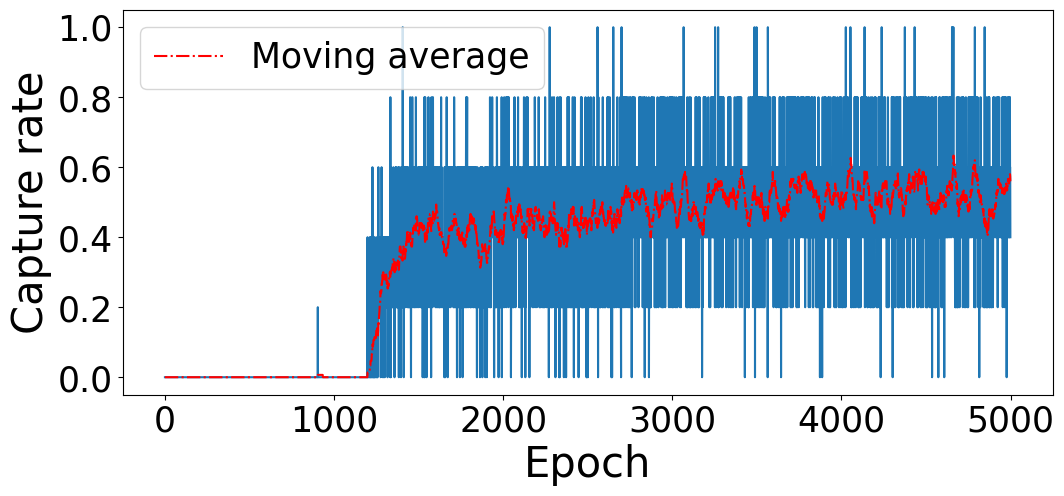}
\label{figure_pursuit_evasion_s_capture_rate}}
\\
\subfloat{
\centering
\includegraphics[width=0.6\linewidth]{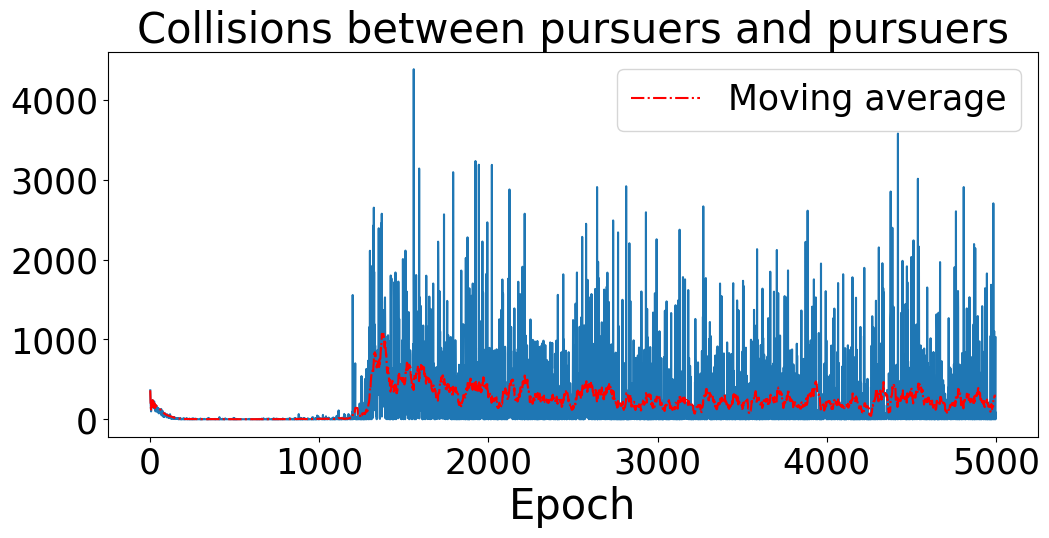}
\label{figure_pursuit_evasion_s_collision}}
\caption{Training performance achieved for Pursuit-Evasion-S.}
\label{fig_pursuit_evasion_s}
\end{figure}

Compared with Pursuit-Evasion-O, more multi-agent conflicts of interest may occur in Pursuit-Evasion-S since more explicit multi-agent coordination is required to capture each evader with four pursuers.
Therefore, we demonstrate the difficulties faced by general MARL in  guaranteeing safe multi-agent coordination with only negative rewards for collisions in Pursuit-Evasion-S.
As shown in Figure \ref{fig_pursuit_evasion_s}, after training for 1000+ epochs, the reward and capture rates improve significantly.
However, more conflicts of interest occur with increasingly efficient capture behavior, as shown in the collision statuses.
The multi-agent collisions do not absolutely vanish with a longer training time, especially in test scenarios, although the costs of collisions are larger than the benefits of capturing an evader in the reward structure (Table \ref{tbl_reward function}).

%


\end{document}